# A novel two-dimensional all-carbon Dirac node-line semimetal

Youjie Wang[1], Qian Gao[1], Zhenpeng Hu[1, *]

1 School of Physics, Nankai University, 94 Weijin Road, Tianjin 300071, China

* zphu@nankai.edu.cn

## Abstract

Carbon allotropes have vast potential in various applications, including superconductivity, energy storage, catalysis, and photoelectric semiconductor devices. Recently, there has been significant research interest in exploring new carbon materials that exhibit unique electronic structures. Here, we propose a novel two-dimensional (2D) carbon allotrope called TCH-SSH-2D, which possesses a Dirac node-line (DNL) semimetallic state. The structure of TCH-SSH-2D is derived from the TCH-type Archimedean polyhedral carbon cluster units, combined with the SSH lattice model, possessing a space group of tetragonal P4/mmm. Using first-principles calculations, we demonstrate that the system is dynamically, thermodynamically, and mechanically stable. It exhibits an energetically favorable structure with no imaginary frequency in the phonon dispersion curves and elastic constants satisfying the Born-Huang stability criterion. Our findings not only contribute to a deeper understanding of the carbon allotrope family but also provide an opportunity to explore unique Dirac states in two-dimensional pure carbon systems.

## 1 Introduction

Carbon is an exceptionally versatile element in the periodic table, capable of forming various hybrid modes of sp, $sp^2$, and $sp^3$. It exhibits the ability to create compounds with other elements and can also be synthesized into a diverse range of carbon allotropes, such as graphene, fullerene, graphene diyne, carbon nanotubes, and carbon foam, which span across different dimensions [1-6]. What makes carbon allotropes even more fascinating is their rich electronic properties and tunable band gaps, which are determined by their unique lattice structures. This allows carbon-based materials to exhibit a wide range of electronic behaviors, encompassing insulators, semiconductors, and even conductors [7-11].

Topological semimetals, as a novel class of quantum materials exhibiting unique properties, are characterized by the intersection and inversion of energy bands near the Fermi level, leading to the emergence of gapless electronic states [12-17]. Based on the degeneracy and spatial distribution of the energy band intersections, topological semimetals can be divided into several categories: Topological Dirac (TD) [15,16,18,19], Topological Weyl (TW) [20-22], and Dirac node-line (DNL) [23-27] semimetals. Extensive research has been conducted on topological materials, with graphene being a prominent example of a TD semimetal featuring stable Dirac points [28-30]. For graphene, the strength of the spin-orbit coupling (SOC) interaction is relatively weak (~$10^{-3}$ meV), and the degeneracy of energy levels is protected by a combination of inversion and time inversion symmetry [31]. It is worth noting that introducing sufficiently strong SOC into graphene can detach the Dirac point, open the gap at the Fermi level, and generate a quantum spin Hall insulating state with conductive edge channels [32]. Unlike the isolated quadruple or double degenerate Dirac points in TD and TW semimetals, DNL semimetals exhibit a unique feature where the linearly dispersed energy planes of the valence and conduction bands near the Fermi level intersect in the momentum space, forming a closed loop [33-34]. Furthermore, in the case of an electrically neutral surface, a distinct semi-filled flat surface state, resembling a drum shape, can emerge. This specific surface state holds great potential for investigating strong electron correlation effects [35]. Close to Dirac point materials, Dirac node-line materials also possess intriguing electronic properties. These distinctive topological band structures give rise to various interesting phenomena, including the semi-integer quantum Hall effect [36], the existence of massless fermions [37], Klein tunneling [38], large magnetoresistance [39], and special optical properties [40]. Currently, DNL semimetals have primarily been predicted in bulk materials, such as antiperovskite $Cu_3(Pd,Zn)N$ [41], $Ca_3P_2$ [42], LaN [43], $PtSn_4$ [44], ZrSiS [45], photonic crystals [46], pure alkali earth metals [47], as well as 3D carbon allotropes [48-49]. In 2015, X. Dai et al. reported the discovery of a pure carbon three-dimensional Mackay Terranes crystal with topological node-lines and 3D Dirac points [48], while the understanding of DNL semimetals in 2D all-carbon systems is still limited. Currently, only a few examples of 2D all-carbon DNL semimetals (H4,4,4-graphyne [50-51], 123-E8Y24-1 [52], and 127-11-84-r568-1 [53]) have been reported with the planar structures of $sp^2$ and $sp^1$ carbon in the literature, while a 2D all-carbon DNL semimetal containing $sp^3$ carbon are still not reported.

Recently, we have employed Archimedean polyhedra as fundamental building blocks and established two connection rules do design carbon materials [54]. Following the same design scheme, we here propose a two-dimensional carbon allotrope that conforms to the Su-Schrieffer-Heeger (SSH) lattice model. Remarkably, this carbon allotrope exhibits the characteristics of a Dirac node-line semimetal. As a unique topological material, the valence and conduction bands of DNL semimetals intersect to form a continuous closed node-line at the Fermi level in momentum space, and this node-line is consistently protected by chiral/mirror/glide mirror symmetries. Our proposed 2D carbon structure holds significant importance in addressing the limitations within the field of 2D topological node-line semimetals.

## 2 Method

All calculations were performed using Density Functional Theory (DFT). Force constants for phonon dispersion and Molecular Dynamics (MD) calculations were conducted using the Vienna Ab-initio Simulation Package (VASP) [55-58], while all other aspects were carried out using the DS-PAW software [59]. The General Gradient-Approximation (GGA) with Perdew-Burke-Ernzerhof (PBE) [60] exchange-correlation functional and Projector-Augmented-Wave (PAW) pseudopotential were employed. The electron wave function was expanded using a plane wave basis set, with a cutoff energy set to 520 eV. The structure was relaxed without symmetry constraints, and the convergence criteria for the total energy and residual forces of each ion were set to $10^{-5}$ eV and 0.01 eV/Å, respectively. The calculation of phonons was carried out using the PHONOPY software package [61]. And a total energy convergence standard was set to $10^{-8}$ eV in the calculations to get force constants. The elastic constants describing the mechanical properties were calculated using the energy-strain method [62-63]. MD simulations were performed adopting a p(3×3×1) supercell using Nose-Hoover thermostat and canonical ensemble (NVT) for a 10 ps duration, employing a time step of 2 fs, at temperatures of 800, 1100, and 1400K.

## 3 Results and discussions

### 3.1 Geometric structure

This carbon allotrope, named TCH-SSH-2D, is derived from Truncated Cubotahedron (TCH) polyhedral structural units and possesses a Su-Schrieffer-Heeger (SSH) lattice structure. TCH is a semiregular polyhedron with 26 faces, 48 vertices, and 72 edges (Figure 1(a)). Figure 1(b) depicts a cubic primitive cell containing a TCH-type 0D carbon cluster. The structure of TCH-SSH-2D is shown in Figures 1(c)-(d), where two TCH units are connected by sharing a common surface (indicated by the purple area) in a planar space. TCH-SSH-2D possesses a square lattice with P4/mmm (No.123) symmetry, and its optimized lattice constant is approximately a=b=6.74 Å. Each unit cell comprises 40 carbon atoms, with a ratio of 3/2 between $sp^2$ hybridized (24 atoms) and $sp^3$ hybridized (16 atoms) carbon atoms. This is the first time that $sp^3$ carbon atoms have been discovered in two-dimensional pure carbon Dirac node-line semimetals. Notably, TCH-SSH-2D exhibits six inequivalent C-C bonds. The shortest bond length measures 1.35 Å, while the longest bond length reaches 1.54 Å. Among these bonds, those with lengths of 1.50, 1.51, 1.52, and 1.54 Å correspond to $sp^3$ hybridization, while the remaining bonds are $sp^2$ hybridized. In Table 1, the bond length information of TCH-SSH-2D was compared to that of graphene ($sp^1$) [64], α-graphyne ($sp^1$, $sp^2$) [65], and diamond ($sp^3$) [66]. The detailed coordinates information for each carbon atom in the structure can be found in Appendix 1.

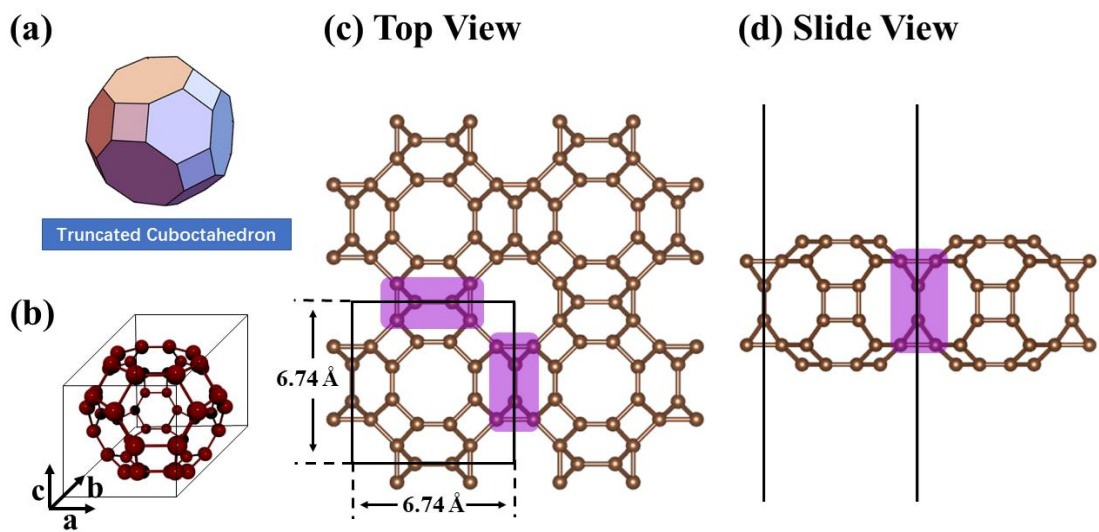

**Fig. 1.** The (a) Truncated cube octahedron (TCH) polyhedron, (b) cubic primitive cell of TCH-type 0D carbon cluster. And the (c) top view, (d) side view of the lattice structure of TCH-SSH-2D.

**Table 1**

The space group, number of carbon atoms in unit cell (N), lattice constants (a / b, Å), and bond length ($d_{C-C}$, Å) for TCH-SSH-2D. For comparison, the corresponding parameters for graphene [64], α-graphyne [65] and diamond [66] are also provided here.

| Structure | Space Group | N | a / b (Å) | $d_{C-C}$ (Å) |
|---|---|---|---|---|
| TCH-SSH-2D | P4/mmm(No.123) | 40 | 6.74 / 6.74 | 1.35  1.46  1.50  1.51  1.52  1.54 |
| Graphene | P6/mmm(No.191) | 2 | 2.47 / 2.47 | 1.43 |
| α-Graphyne | P6/mmm(No.191) | 8 | 6.96 / 6.96 | 1.23  1.40 |
| Diamond | Fd3m(No.227) | 8 | 3.57 / 3.57 | 1.55 |

## 3.2 Dynamics, thermodynamics, and mechanical stability

To confirm the structural stability of TCH-SSH-2D, we investigated its dynamics, thermodynamics, and mechanical stability. We obtained the phonon dispersion curves along the highly symmetric direction of the first Brillouin zone, as illustrated in Figure 2(a). There is no imaginary frequency, and the maximum frequency is 48.96 THz, which indicates the dynamic stability of this newly discovered two-dimensional carbon allotrope.

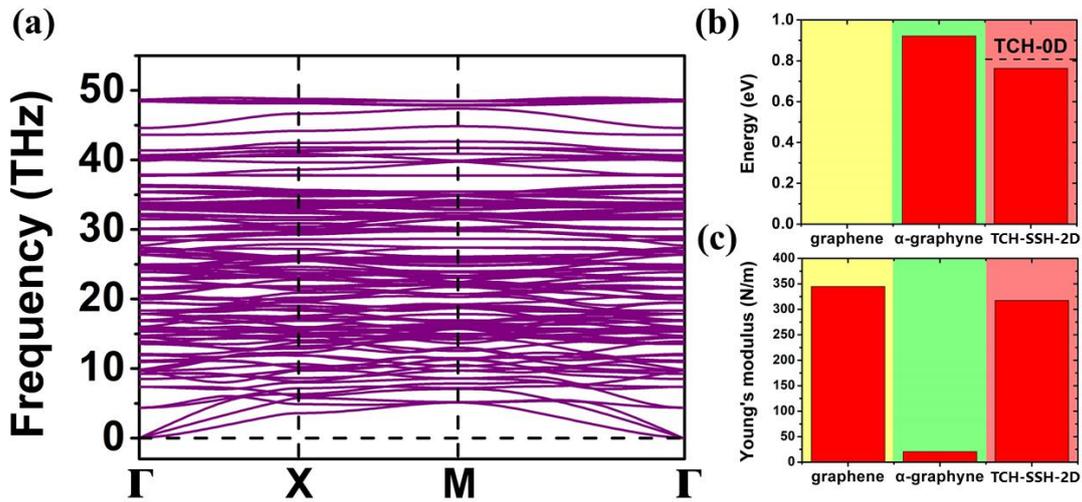

**Fig. 2.** (a) The phonon dispersion curves along the highly symmetric direction in the first Brillouin region of TCH-SSH-2D, (b) energy per atom of carbon allotropes, (c) Young's modulus of carbon allotropes. The energy of graphene is set to 0 as a reference.

Figure 2(b) presents the single atom energies of graphene, α-graphyne, and TCH-SSH-2D. To facilitate a clear comparison, we have set the energy per atom of graphene as the reference point (0). In comparison to graphene, TCH-SSH-2D exhibits a higher energy, with an increase of 0.763 eV/atom. However, when compared to α-graphyne and TCH-based 0D carbon cluster, TCH-SSH-2D is more energetically advantageous, suggesting its relative stability. To further confirm the thermodynamic stability of TCH-SSH-2D, we constructed a p(3×3×1) supercell and performed ab initio molecular dynamics (AIMD) simulations using the Nose-Hoover thermostat and a 2 fs time-step canonical ensemble (NVT). We examined the temperature and total energy fluctuations of the system over time at temperatures of 800 K, 1100 K, and 1400 K. As depicted in Figures 3(a)-(c), during the 10 ps MD simulation, TCH-SSH-2D exhibits minimal fluctuations in total energy and temperature, indicating its robust thermodynamic stability. With every 300 K increment in temperature, the average energy of the system increases by approximately 10 eV. Figures 3(d)-(f) display corresponding snapshots at temperatures of 800 K, 1100 K, and 1400 K, respectively. The geometric network of TCH-SSH-2D at specific time step in the MD simulation process demonstrates the preservation of its structure without any noticeable reconstruction.

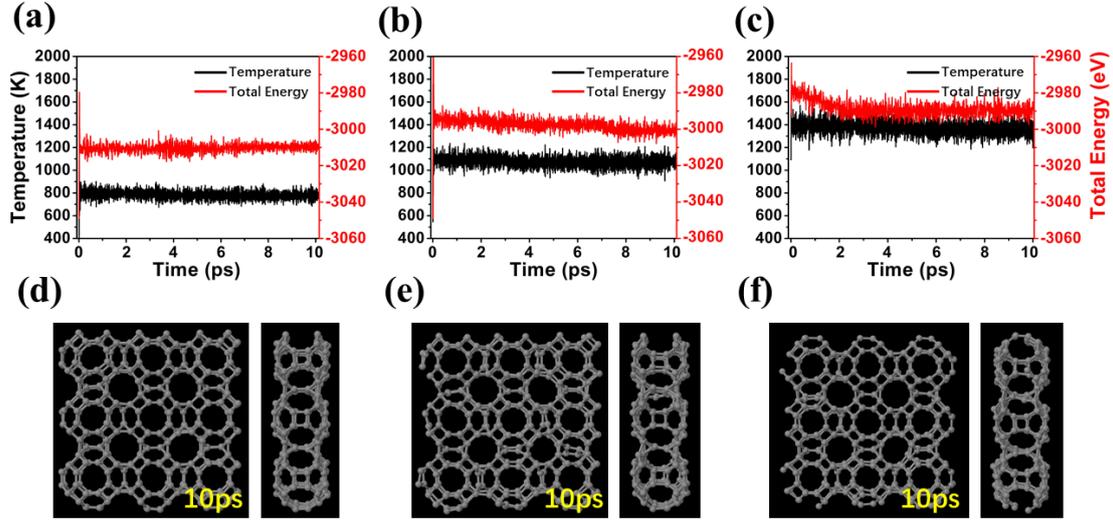

**Fig. 3.** (a) - (c) The temperature and total energy fluctuations of TCH-SSH-2D as a function of simulation time at temperatures of 800 K, 1100 K, and 1400 K, respectively. (d) - (f) The corresponding snapshots of the systems with p(3×3×1) supercell structures at 10ps in the MD simulation at temperatures of 800 K, 1100 K, and 1400 K, respectively.

The mechanical properties of TCH-SSH-2D were investigated and compared with those of graphene and α-graphyne to assess the stability and mechanical behavior of this carbon material for potential applications. The square structure of TCH-SSH-2D has three independent elastic constants: $C_{11}$, $C_{12}$, and $C_{66}$. The in-plane elastic constants were calculated by fitting the strain energy curve under small strain, obtaining the following values: $C_{11} = C_{22} = 354.15$ N/m, $C_{12} = 61.91$ N/m, and $C_{66} = 128.25$ N/m. Based on the Born-Huang criteria, which states that $C_{11} > 0$, $C_{66} > 0$, and $|C_{11}| > |C_{12}|$, all the elastic constants of TCH-SSH-2D satisfy the criteria for mechanical stability. Figure 2(c) presents the histogram of Young's modulus, revealing that the Young's modulus of TCH-SSH-2D is significantly higher than that of α-graphyne (20.67 N/m). The maximum and minimum values of Young's modulus for TCH-SSH-2D are 343.321 N/m and 317.355 N/m, respectively, indicating no significant anisotropy. Additional information regarding the mechanical properties can be found in Table 2.

**Table 2**

The independent elastic constants ($C_{ij}$, N/m), maximum/minimum value of Young's modulus ($Y_s$, N/m) /Shear modulus (S, N/m) /Poisson's ratio (P, N/m), and total energy per atom ($E_{tot}$, eV/atom) of TCH-SSH-2D, graphene [64, 65, 67, 68] and α-graphyne [68].

| Structure | $C_{11}/C_{22}$ | $C_{12}/C_{21}$ | $C_{66}$ | Ys-min\|max | S-min\|max | P-min\|max | $E_{tot}$ |
| --- | --- | --- | --- | --- | --- | --- | --- |
| TCH-SSH-2D | 354.15 | 61.91 | 128.25 | 317.36 \| 343.32 | 128.25 \| 146.12 | 0.175 \| 0.2 | -8.46 |
| Graphene | 355.73 | 62.25 | 146.74 | 344.83 \| 344.83 | 146.73 \| 146.73 | 0.175 \| 0.175 | -9.22 |
| α-Graphyne | 95.10 | 84.13 | 5.49 | 20.68 \| 20.68 | 5.49 \| 5.49 | 0.885 \| 0.885 | -8.30 |

### 3.3 Electronic characteristics

The 2D band structure of TCH-SSH-2D, depicted in Figure 4(a), revealed two non-equivalent intersections between the lowest conduction band (LCB) and highest valence band (HVB) around the Γ point along the M-Γ and Γ-X paths. This observation sparked our interest and prompted us to conduct first-principles calculations to obtain the 3D band structure of the entire first Brillouin zone, as shown in Figure 4(d). It became evident that the LCB and HVB of TCH-SSH-2D intersect and undergo bands inversion near the Fermi level, resulting in the formation of a node-line centered around the Γ point. To emphasize the node-line, we depicted a Fermi circular loop with a radius of approximately 0.2 (2π/a) using black solid lines in Figures 4(d)-(e), representing the Dirac node-line. Based on the obtained electronic band structure, the Fermi velocities inside and outside the Dirac node loop are calculated to be 1.68 and $5.09 \times 10^5$ m/s in all directions. The Fermi velocities of TCH-SSH-2D exhibit isotropy and share the same order of magnitude with that of graphene (~$9.0 \times 10^5$ m/s) [69]. The symmetrical constraint imposed by the crystal structure of TCH-SSH-2D prevents the node-lines from being disrupted or damaged by small disturbances, ensuring the robust existence and protection of the node-line feature. Upon analyzing the orbital-resolved electron band structure depicted in Figure 4(b), we observed that both energy bands near the Fermi level are predominantly contributed by the p orbital. The size of the icon bubble represents the contribution of the corresponding atomic orbitals to the energy bands. Additionally, the density of states projection in Figure 4(c) confirms that the density of states near the Fermi level is primarily attributed to the p orbitals. Our work further extends the concept of DNL from 3D to 2D systems, introducing a novel two-dimensional Dirac node-line carbon allotrope. This discovery provides a new potential candidate for carbon-based high-speed electronic devices.

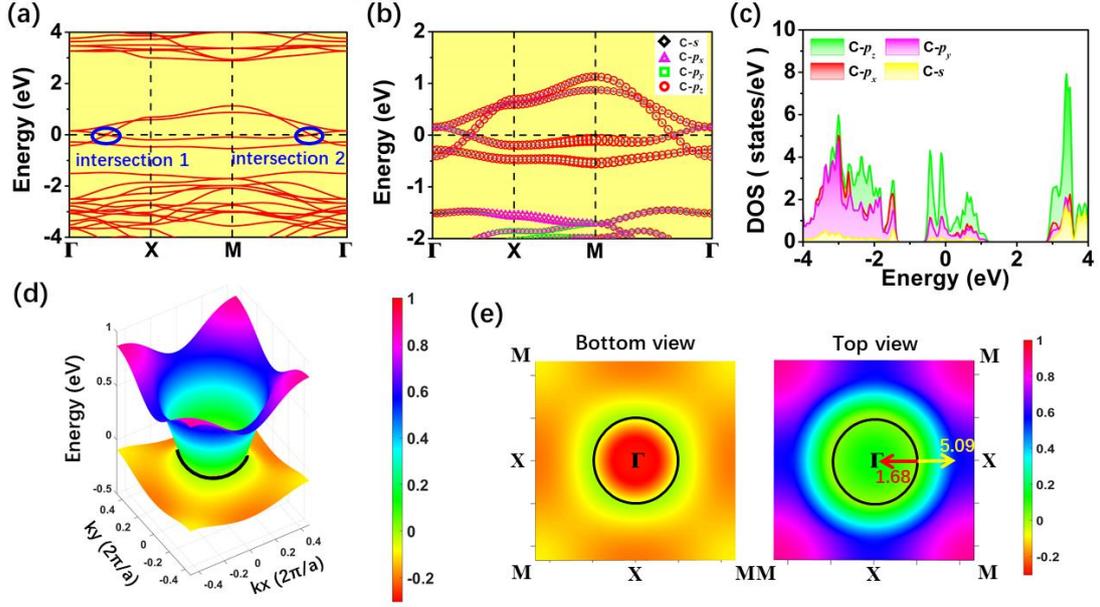

**Fig. 4.** The (a) band structure along high symmetry direction of the first Brillouin zone, (b) atomic orbital projected band (AOPB), (c) projected density of states (PDOS), (d) 3D band structure of the first Brillouin zone, (e) 3D band structure: bottom view, top view of TCH-SSH-2D. The size of the icon bubble in (b) indicates the contribution of the corresponding atomic orbitals to the energy bands. The black solid lines in (d)-(e) represent the node-line. The Fermi level is set to 0 eV.

## 4 Conclusion

In summary, we proposed and validated a two-dimensional pure carbon Dirac node-line semimetal called TCH-SSH-2D through first principles calculations. This 2D structure is constructed based on the Su-Schrieffer-Heeger (SSH) lattice model, utilizing Truncated Cubotahedron (TCH) polyhedra as building units. TCH-SSH-2D has a square lattice with P4/mmm (No. 123) symmetry, and its optimized lattice constant is a=b=6.74 Å. Its unit cell contains 40 carbon atoms, with 24 atoms in the $sp^2$ hybrid state and the remaining 16 atoms in the $sp^3$ hybrid state. This is the first discovery of a Dirac node-line semimetal containing $sp^3$ carbon atoms in the two-dimensional carbon system. First principles calculations indicated that TCH-SSH-2D has dynamics, thermodynamics, and mechanical stability. The node-line structure of TCH-SSH-2D is a circular loop centered at the Γ point, with a radius of 0.2 (2π/a). The Fermi velocities inside and outside the node-line loop are 1.68 and $5.09×10^5$ m/s, respectively, in all directions. The energy bands near the Fermi level are predominantly contributed by the p orbital. The existence of node-line structure positions TCH-SSH-2D as a highly promising material for high-speed electronic devices.

## Acknowledgements

This work was supported by the National Natural Science Foundation of China (Grant Nos. 12134019 and 21773124), the Fundamental Research Funds for the Central Universities Nankai University (No. 010-63233001, 63213042, 63221346, and ZB22000103), and the Supercomputing Center of Nankai University (NKSC). Y. Wang was partially supported by the

## Appendix 1

### TCH-SSH-2D

**The lattice constants**

| | | |
|---|---|---|
| 6.7362519891991761 | -0.0000000000000000 | 0.0000000000000000 |
| -0.0000000000000000 | 6.7362519891991761 | 0.0000000000000000 |
| 0.0000000000000000 | -0.0000000000000000 | 17.6460418126742979 |

**The fractional coordinates for each carbon atom**

| | X | Y | Z |
|---|---|---|---|
| C1 | 0.1129684326065099 | 0.2750681812343467 | 0.3953767275957031 |
| C2 | 0.8870315523934851 | 0.7249318187656536 | 0.6046232424042945 |
| C3 | 0.8870315523934851 | 0.7249318187656536 | 0.3953767275957031 |
| C4 | 0.1129684326065099 | 0.2750681812343467 | 0.6046232424042945 |
| C5 | 0.7249318187656536 | 0.1129684326065099 | 0.3953767275957031 |
| C6 | 0.2750681812343467 | 0.8870315523934852 | 0.6046232424042945 |
| C7 | 0.2750681812343467 | 0.8870315523934851 | 0.3953767275957031 |
| C8 | 0.7249318187656536 | 0.1129684326065099 | 0.6046232424042945 |
| C9 | 0.8870315523934851 | 0.2750681812343467 | 0.6046232424042945 |
| C10 | 0.1129684326065099 | 0.7249318187656536 | 0.3953767275957031 |
| C11 | 0.1129684326065099 | 0.7249318187656536 | 0.6046232424042945 |
| C12 | 0.8870315523934851 | 0.2750681812343467 | 0.3953767275957031 |
| C13 | 0.2750681812343467 | 0.1129684326065099 | 0.6046232424042945 |
| C14 | 0.7249318187656536 | 0.8870315523934852 | 0.3953767275957031 |
| C15 | 0.7249318187656536 | 0.8870315523934852 | 0.6046232424042945 |
| C16 | 0.2750681812343467 | 0.1129684326065099 | 0.3953767275957031 |
| C17 | -0.0000000000000000 | 0.3854267172241525 | 0.4570913533045587 |
| C18 | -0.0000000000000000 | 0.6145733127758498 | 0.5429086766954438 |
| C19 | -0.0000000000000000 | 0.6145733127758498 | 0.4570913533045587 |
| C20 | -0.0000000000000000 | 0.3854267172241525 | 0.5429086766954438 |
| C21 | 0.6145733127758498 | 0.0000000000000000 | 0.4570913533045587 |
| C22 | 0.3854267172241525 | 0.0000000000000000 | 0.5429086766954438 |
| C23 | 0.3854267172241525 | 0.0000000000000000 | 0.4570913533045587 |
| C24 | 0.6145733127758498 | 0.0000000000000000 | 0.5429086766954438 |
| C25 | 0.2460049589421658 | 0.3997580652366269 | 0.3465473044399731 |
| C26 | 0.7539950260578363 | 0.6002419347633732 | 0.3465473044399731 |
| C27 | 0.6002419347633732 | 0.2460049589421658 | 0.3465473044399731 |
| C28 | 0.3997580652366269 | 0.7539950260578363 | 0.3465473044399731 |
| C29 | 0.7539950260578363 | 0.3997580652366269 | 0.3465473044399731 |
| C30 | 0.2460049589421658 | 0.6002419347633732 | 0.3465473044399731 |
| C31 | 0.3997580652366269 | 0.2460049589421658 | 0.3465473044399731 |
| C32 | 0.6002419347633732 | 0.7539950260578363 | 0.3465473044399731 |
| C33 | 0.2460049589421658 | 0.3997580652366269 | 0.6534527245600259 |
| C34 | 0.7539950260578363 | 0.6002419347633732 | 0.6534527245600259 |
| C35 | 0.6002419347633732 | 0.2460049589421658 | 0.6534527245600259 |
| C36 | 0.3997580652366269 | 0.7539950260578363 | 0.6534527245600259 |
| C37 | 0.7539950260578363 | 0.3997580652366269 | 0.6534527245600259 |
| C38 | 0.2460049589421658 | 0.6002419347633732 | 0.6534527245600259 |
| C39 | 0.3997580652366269 | 0.2460049589421658 | 0.6534527245600259 |
| C40 | 0.6002419347633732 | 0.7539950260578363 | 0.6534527245600259 |